\documentclass[useAMS,usegraphicx]{mn2e}
\usepackage{times}
\usepackage{rotate}
\usepackage{amssymb}
\usepackage{multirow}
\title[X-ray selected AGN in groups at redshifts $z \approx 1$]
{X-ray selected AGN in groups at redshifts $z \approx 1$}
\author[Georgakakis et al. ] {A. Georgakakis$^{1,2}$,   Brian
  F. Gerke$^3$, K. Nandra$^{1}$, E. S. Laird$^{1}$, 
  A. L. Coil$^{4}$, M. C. Cooper$^4$,\\\\   
  {\rm \LARGE J. A. Newman$^5$}\\\\
  $^1$Astrophysics Group, Blackett Laboratory, Imperial College, Prince
  Consort Rd , London SW7 2BZ, UK\\ 
  $^2$National Observatory of Athens, I. Metaxa \& V. Paulou,
  Athens 15236, Greece\\ 
  $^3$Kavli Institute for Particle Astrophysics and Cosmology,
  Stanford Linear Accelerator Center, 25 Sand Hill Rd. MS 29, Menlo
  Park, CA 94725, USA\\ 
  $^4$Steward Observatory, University of Arizona, 933 N. Cherry Ave.,
  Tucson, AZ 85721-0065, USA\\
  $^5$University of Pittsburgh, Physics \& Astronomy Department,
  3941 O'Hara Street Pittsburgh, PA, 15260, USA
}
\begin{document}
\maketitle

\begin{abstract} We explore  the role of the group  environment in the
evolution of  AGN at the  redshift interval $0.7<z<1.4$,  by combining
deep Chandra observations with extensive optical spectroscopy from the
All-wavelength Extended Groth strip International Survey (AEGIS).  The
sample consists of 3902 optical  sources and 71 X-ray AGN. Compared to
the overall  optically selected galaxy population, X-ray  AGN are more
frequently  found in  groups at  the 99\%  confidence level.   This is
partly  because AGN  are hosted  by red  luminous galaxies,  which are
known to reside, on average, in dense environments.  Relative to these
sources, the excess of X-ray AGN  in groups is significant at the 91\%
level only.  Restricting the  sample to $0.7<z<0.9$ and $M_B<-20$\,mag
in order  to control systematics we  find that X-ray  AGN represent
$(4.7\pm1.6)$ and $(4.5\pm1.0)$\% of  the optical galaxy population in
groups and  in the field  respectively.  These numbers  are consistent
with the AGN fraction in  low redshift clusters, groups and the field.
The  results  above, although  affected  by  small number  statistics,
suggest that X-ray  AGN are spread over a  range of environments, from
groups to the field, once  the properties of their hosts (e.g. colour,
luminosity)  are accounted  for.   There is  also tentative  evidence,
significant  at the  98\% level,  that the  field produces  more X-ray
luminous  AGN compared  to groups,  extending similar  results  at low
redshift to $z\approx1$.  This trend may be because of either cold gas
availability or the nature of the interactions occurring in the denser
group environment (i.e.  prolonged tidal encounters).
\end{abstract}
\begin{keywords}  
  Surveys  --  galaxies:  active  --  galaxies:
  structure -- cosmology: large scale structure of the Universe
\end{keywords}

\section{Introduction}\label{sec_intro}

The environment of Active  Galactic Nuclei (AGN) holds important clues
on the physical mechanism(s) responsible for the accretion of material
on the  central supermassive black hole  (SBH).  Numerical simulations
have  shown that an  efficient way  of channeling  gas to  the nuclear
galaxy regions and in the  vicinity of the SBH are galaxy interactions
and  mergers  (e.g.  Barnes  \&  Hernquist  1992;  Mihos \&  Hernquist
1996). As a result, many models for the formation and the cosmological
evolution of AGN assume that the  main growth phase of the central SBH
in galaxies occurs in major  mergers (e.g. Kauffmann \& Haehnelt 2000;
Di Matteo, Springel  \& Hernquist; Hopkins et al.   2005; Bower et al.
2006;  Croton  et  al.   2006).   Recent  observations  of  the  local
environment of  luminous QSOs supports  this scenario.  The  excess of
optical  neighbours  around $z<0.4$  QSOs  on  small scales  ($\approx
0.1-0.5$\,Mpc) compared to $L_{\star}$  galaxies (Serber et al.  2006)
and  the higher  fraction of  QSO pairs  with  separations $<0.1$\,Mpc
compared to  the expectation from large scales  ($>3$\,Mpc; Hennawi et
al.  2006; Myers et al.   2007), both suggest that galaxy interactions
may play  an important role in  the evolution of QSOs.   This does not
appear to be  the case however, for the more  common but less powerful
narrow-line AGN (Seyfert, Low Ionization Nuclear Emission Line Regions
-- LINERs) selected in the Sloan Digital Sky Survey (SDSS) at $z<0.3$.
The small  scale environment  of these sources  ($<$0.1\,Mpc) suggests
that galaxy interactions alone cannot explain the observed activity in
the bulk of  this population (Li et al.   2006).  Modeling work indeed
predicts that  low luminosity  AGN, such as  Seyferts and  LINERs, may
accrete   cold  gas   stochastically   by  mechanisms   such  as   bar
instabilities or  tidal disruptions rather than  major interactions or
mergers (Hopkins \& Hernquist 2006).

On scales  much larger than  0.5\,Mpc, the clustering of  AGN measures
the mass  of their  host dark matter  halo and  how it relates  to the
observed  nuclear activity.  At  these scales,  an increasing  body of
evidence underlines the potential significance in the evolution of AGN
of the group environment.  Early  studies for example, have shown that
low redshift  luminous QSOs are associated  with moderate enhancements
in the distribution  of galaxies and avoid rich  clusters (Hartwick \&
Schade 1990; Bahcall  \& Chokshi 1991; Fisher et  al.  1996; McLure \&
Dunlop 2001). Moreover, the evolution of the large scale clustering of
broad line QSOs  in the redshift range $0.5-2.5$  is consistent with a
parent dark matter halo of  almost constant mass in the range $10^{12}
- 10^{13}\, \, h^{-1} \rm M_{\odot}$  (Porciani et al.  2004; Croom et
al.  2005; Hopkins et al.  2007; da Angela et al. 2007; Mountrichas et
al.  2008;  but see  Coil et  al.  2007), close  to the  threshold for
galaxy groups ($6 \times 10^{12} \, \, h^{-1} \rm M_{\odot}$ for DEEP2
groups with velocity  dispersion $\rm \sigma_v>200\, km\,s^{-1}$; Coil
et  al 2006) The  more common  narrow emission-line  AGN in  the local
Universe  ($z\approx0.1$) are  hosted by  massive  early-type galaxies
(Kauffmann  et al.  2003),  which are  known to  be more  clustered on
Mpc-scales than  the overall galaxy population (Zehavi  et al.  2005).
A luminosity  dependence of the  clustering has also been  found, with
the most  luminous narrow-line AGN  residing in lower  density regions
compared to  less luminous ones (Kauffmann  et al.  2004;  Wake et al.
2004; Constantin \& Vogeley  2006).  Popesso \& Biviano (2006) further
showed that  not only the luminosity but  also the number of  AGN is a
strong  function  of local  density.   In  a  sample of  low  redshift
clusters,   they  found   that   the  fraction   of  narrow-line   AGN
anti-correlates  with cluster  velocity dispersion,  increasing toward
galaxy groups.

At  higher  redshift,  $z\approx1$,  X-ray surveys  provide  the  most
efficient and least biased way  of finding active SBH to explore their
demographics.  The large scale structure  of the X-ray selected AGN at
$z\approx1$  is usually determined  using the  angular or  the spatial
auto-correlation functions  (e.g. Yang et al.  2004;  Basilakos et al.
2004, 2005; Gilli et al.   2005; Miyaji et al.  2007).  Although there
is  substantial scatter  in the  results from  these  studies, largely
because  of  observational  uncertainties  (Miyaji et  al.   2007)  or
luminosity  dependent clustering  (Plionis et  al. 2007),  the general
consensus is that X-ray selected  AGN are hosted by dark matter haloes
with  mass of about  $10^{13} -  10^{14} \,  \, h^{-1}  \rm M_{\odot}$
(Miyaji et  al. 2007), i.e.  in  the range expected  for galaxy groups
(e.g.  Zandivarez, Merchan, \& Padilla 2003).

Why AGN  show a  preference for the  group scale environment  is still
unclear.   One  possibility is  that  galaxy  interactions, which  are
believed  to be one  of the  AGN triggers,  are suppressed  in massive
clusters, while  they proceed more  efficiently in the  lower velocity
dispersion groups  of galaxies.  Alternatively,  it might be  that the
necessary ingredients  for AGN activity, sufficient  cold gas supplies
and a  SBH, are  often found in  moderate density  environments.  More
massive galaxies,  which are likely  to host larger black  holes (e.g.
Ferrarese \& Merritt 2000; Gebhardt et al.  2000), are associated with
high  density  regions  (e.g.   Hogg  et  al.   2003;  Cooper  et  al.
2006).  Moreover, cold  gas, the  fuel of  AGN, is  likely to  be more
abundant in  moderate density environments, while it  is depleted more
efficiently in  high density regions,  such as clusters.   Despite the
potential significance  of galaxy groups, there are  still few studies
that attempt  to directly  constrain the incidence  of active  SBHs in
these moderate  density environments.  Shen  et al. (2007)  found that
the fraction of AGN (both X-ray and optically selected) in a sample of
eight  poor groups  at $z  \approx 0.06$  is consistent  with  that of
clusters or  the field (Martini  et al. 2007).  Although  small number
statistics are a  issue in these studies, the  finding above is against
scenarios where  the group environment promotes  AGN activity, because
of e.g.  the higher efficiency of galaxy-galaxy mergers.

At higher redshift, $z\approx1$, close  to the peak of the AGN density
in the  Universe (e.g.   Barger et al.  2005; Hasinger et  al.  2005),
there  is still  no information  on  the association  between AGN  and
groups.   This is  mainly  because  it is  hard  to identify  moderate
overdensities  at high  redshift, as  it requires  large spectroscopic
surveys  of  optically faint  galaxies.   The All-wavelength  Extended
Groth-strip  International  Survey (AEGIS;  Davis  et  al.  2007)  has
extensive optical spectroscopy that can  be used to identify groups at
$z\approx1$  and deep  {\it Chandra}  observations, which  provide the
most efficient way for finding  AGN, particularly at high redshift. In
this paper we combine the  AEGIS group and X-ray catalogues to explore
the   fraction   of  X-ray   selected   AGN   associated  with   these
overdensities.  We adopt $H_{0} = \rm 70 \, km \, s^{-1} \, Mpc^{-1}$,
$\Omega_{M} = 0.3$ and $\Omega_\Lambda = 0.7$.

\section{Data and sample selection}
 
The X-ray  data are from  the {\it Chandra}  survey of the  AEGIS. The
observations  consist  of  8  ACIS-I  pointings,  each  with  a  total
integration  time  of  about  200\,ks  split in  at  least  3  shorter
exposures obtained  at different  epochs.  The data  reduction, source
detection and flux estimation are  described in detail by Laird et al.
(in preparation) and  are based on methods presented  by Nandra et al.
(2005).  The  limiting flux in the  0.5-2 and 2-10\,keV  band are $1.1
\times  10^{-16}$,  $\rm 8.2  \times  10^{-16}  \,  erg \,  s^{-1}  \,
cm^{-2}$, respectively.  The X-ray catalogue comprises a total of 1318
sources over  $\rm 0.63\,deg^{2}$  to a Poisson  detection probability
threshold of $4 \times 10^{-6}$. For the optical identification we use
the  DEEP2  photometric catalogues  and  the  Likelihood Ratio  ($LR$)
method (e.g.  Brusa et  al. 2007;  Laird et al.   in prep.).   In this
paper  optical counterparts  are selected  to have  $LR>1$.   For this
cutoff,  there   are  857   optical  identifications  and   the  false
counterpart rate is estimated to be 4.4 per cent.

The AEGIS  is one of  the four fields  targeted by the  DEEP2 redshift
survey. This spectroscopic program uses the DEIMOS spectrograph (Faber
et al.  2003)  on the 10\,m Keck-II telescope  to obtain redshifts for
galaxies  to $R_{AB}  = 24.1$\,mag.   The observational  setup  uses a
moderately high resolution  grating ($R\approx5000$), which provides a
velocity accuracy of $\rm 30\,km\,s^{-1}$ and a wavelength coverage of
6500--9100\,\AA.   This spectral window  allows the  identification of
the strong [O\,II] doublet 3727\AA\, emission line to $z<1.4$.  We use
DEEP2  galaxies with  redshift  determinations secure  at the  $>90\%$
confidence level (quality flag $Q \ge 3$; Davis et al. 2007).

\section{The Group Catalog}

The   AEGIS   group  catalogue   has   been   constructed  using   the
Voronoi-Delaunay  group  finding  method  described by  Gerke  et  al.
(2005). This algorithm  has been optimised and tested  for the typical
DEEP2 redshift survey field,  in which spectroscopic targets have been
pre-selected based on their optical  colour to limit the sample in the
redshift interval  $0.7<z<1.4$.  This is  unlike the AEGIS  field, for
which  no  colour pre-selection  has  been  applied for  spectroscopic
follow ups  and the sampling rate  of the galaxy  population is higher
than the  rest of the DEEP2  survey fields.  The Gerke  et al.  (2005)
group  finding method  has been  applied  on the  sub-sample of  AEGIS
galaxies  that  fulfill   the  DEEP2  spectroscopic  survey  selection
criteria (i.e.  colour pre-selection, sampling rate).  This includes a
total of  3902 optical sources  in the redshift range  $0.7<z<1.4$, of
which 1125 belong  to groups, and 71 X-ray AGN  over the same redshift
interval.

Gerke et  al.  (2005) estimate  that about 79  per cent of  real group
members are assigned to a  group.  The interloper fraction among group
galaxies is  46 per  cent, while the  field sample is  contaminated by
group  members at  the 6  per cent  level only.   Correctly classified
galaxies therefore dominate both  the group and field populations. The
group catalogue of Gerke et al.  (2005) has been designed to reproduce
the distribution of groups as  a function of velocity dispersion, $\rm
\sigma_v$, and  redshift, for $\rm  \sigma_v>350km\,s^{-1}$.  However,
in  this paper we  are interested  in the  properties of  group member
galaxies i.e.,  the fraction of  X-ray AGN in this  population, rather
than the group properties.  Therefore  the large number of groups with
velocity dispersions lower than the limit above are also considered in
the  analysis.   Although  the   uncertainty  of  the  estimated  $\rm
\sigma_v$  is  likely  to  be  large,  particularly  at  low  velocity
dispersions,  because   of  the   discrete  sampling  of   the  galaxy
population,  the  group sample  can  still be  used  to  explore in  a
statistical way  the properties of  the member galaxies.   Groups with
$\rm \sigma_v \la 100\, km  \, s^{-1}$ may suffer higher contamination
(e.g.  Gerke  et al.  2007), but  we nevertheless choose  not to apply
any velocity dispersion cutoff to  the group catalogue used here.  Our
analysis   guarantees  that  this   has  a   minimal  impact   on  the
results. This is  because the incidence of group  galaxy members among
X-ray  sources  is  compared  with  that  of  the  optically  selected
galaxies.  Any contamination  affecting low velocity dispersion groups
cancels out in this differential test.  By including all groups in the
analysis we  also improve the statistical reliability  of the results.
Nevertheless,  restricting the  group galaxy  sample to  those sources
whose host groups have $\rm \sigma_v  \ga 100\, km \, s^{-1}$ does not
alter  our  conclusions. It  is  important  to  note that  the  groups
identified  here have  moderate virial  masses in  the  range $M_{vir}
\approx 5 \times  10^{12} - 5 \times 10^{13}  \, M_{\odot}$, with very
few  groups  having  $M_{vir}  > 10^{14}\,M_{\odot}$.   Therefore  the
results presented here do not apply to clusters.  In the next sections
the  group population  is referred  to those  galaxies that  have been
identified as members  of group. The remaining galaxies  are the field
population.

\begin{table} 
\caption{Fraction of galaxies and X-ray AGN in groups}\label{tab_xdet} 
\begin{center} 
\scriptsize
 \begin{tabular}{l c c c }
\hline 
Sample & $N_{TOT}$ &  $N_{G}$ &  $f_G$ \\
       &          &          &  ($\%$)  \\
  (1)  &  (2)     &    (3)   &  (4)   \\
\hline

X-ray sources  & 71 & 30 & $42.3\pm5.4$ \\

optical galaxies  & 3902 & 1125 & $28.8\pm0.7$ \\

optical galaxies  (matched $M_B$/$U-B$)$^{1}$   &  -- & -- & $33.8\pm5.6$\\

\hline
\end{tabular} 

\begin{list}{}{}
\item 

The columns are:  (1): Sample definition; (2): $N_{TOT}$  is the total
number  of galaxies  in the  sample; (3):  $N_{G}$ corresponds  to the
number  of galaxies  that are  group  members; (4):  $f_G$ is  defined
$f_G=N_{G}/N_{TOT}$.  The errors for  the first two rows are estimated
assuming binomial statistics. For the third row see note below.

\item $^{1}$The 3rd row corresponds  to the sample of optical galaxies
with $M_B$ and $U-B$ distributions  matched to those of X-ray selected
AGN. This sample is not unique and therefore $N_{TOT}$ and $N_{G}$ are
not well defined quantities and  are not listed. For that sample $f_G$
can  only  be  estimated  through  the simulations  described  in  the
text. We list the median of the $f_G$ distribution determined from the
trials described  in the text. The  errors correspond to  the 16th and
84th percentile of the distribution, i.e. the 68\% confidence level.

\end{list}

\end{center}
\end{table}

\begin{figure*}
\begin{center}
 \rotatebox{0}{\includegraphics[height=0.9\columnwidth]{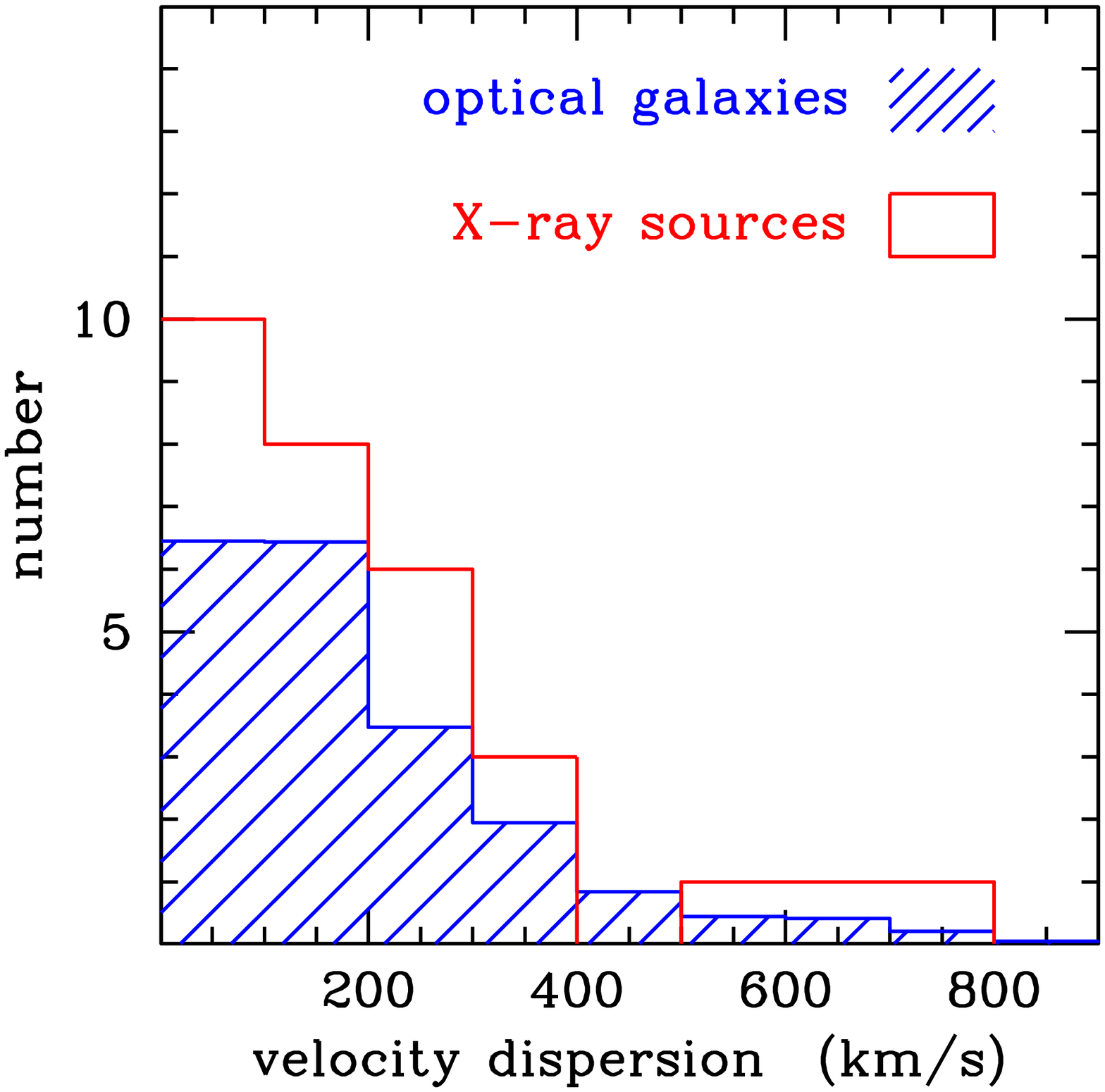}}
 \rotatebox{0}{\includegraphics[height=0.9\columnwidth]{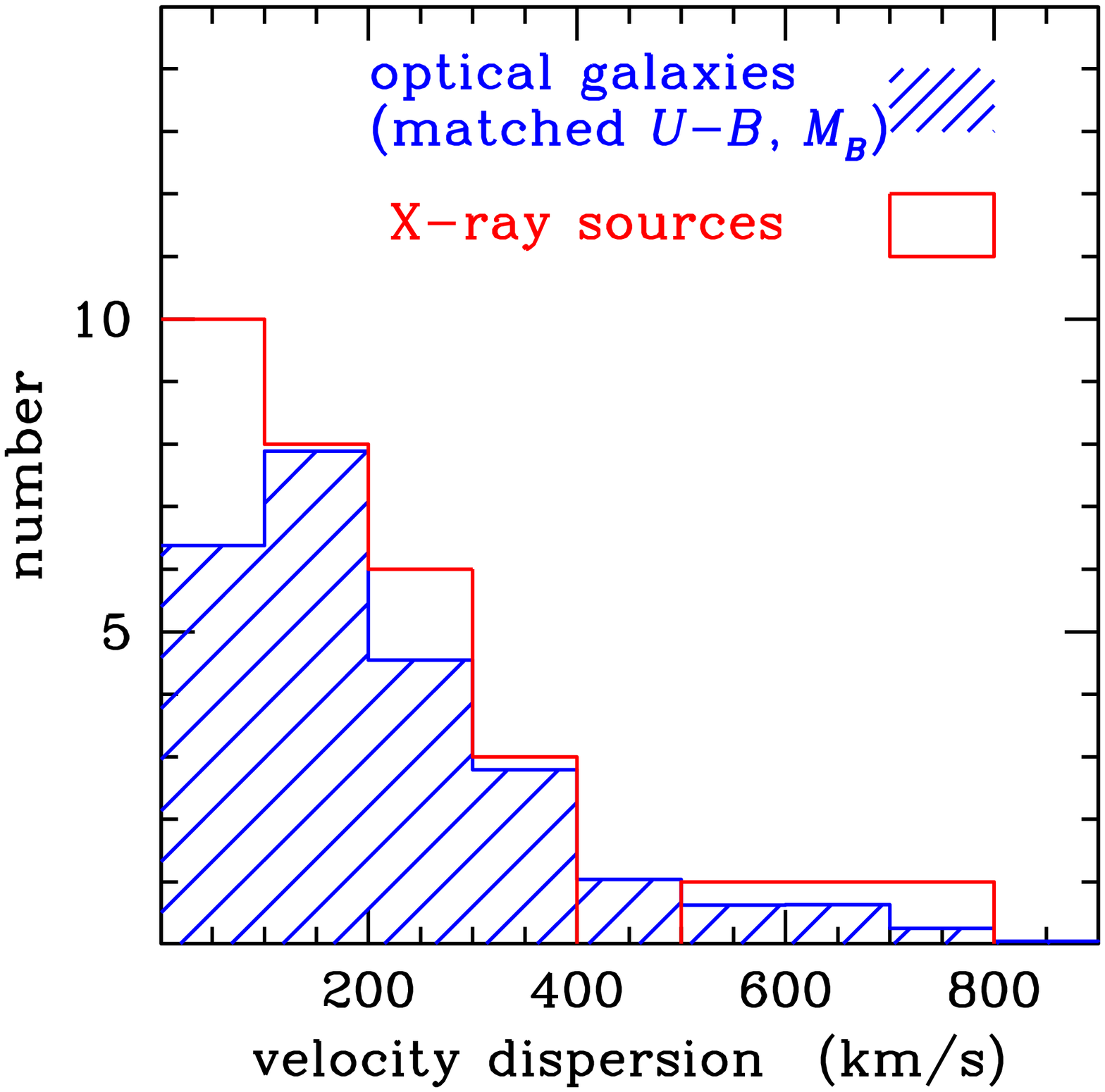}}
\end{center}
\caption{  Group velocity  dispersion histogram.   In both  panels the
open (red) histogram is for  X-ray selected AGN associated with groups
in the  AEGIS survey at $z\approx  1$.  In the left  panel the hatched
(blue)  histogram is  estimated  by resampling  the overall  optically
selected galaxy population as described in the text. X-ray sources are
more frequently found in groups at the 99 per cent significance level.
In  the right  panel  the  hatched (blue)  histogram  is estimated  by
resampling  the optical galaxy  population and  by requiring  that the
$U-B$ and $M_B$ distributions of  the random samples in each trial are
the same  as those of X-ray  sources.  In this case  X-ray sources are
more  frequently found  in  groups  at the  91  per cent  significance
level.}\label{fig_vdisp}
\end{figure*}

\section{Results}\label{results}

\subsection{Comparison of X-ray AGN with optically selected galaxies}

In Table  1 a large  fraction of the  X-ray selected AGN are  found in
groups, about  42 per cent  (30/71).  For comparison, the  fraction of
group members among  optical galaxies is 29 per  cent (1125/3902). The
significance  of the excess  is assessed  by randomly  resampling with
replacement the optical galaxy population to construct subsamples with
size  equal that of  the X-ray  source catalogue.   In each  trial the
fraction  of  group  galaxies  is  registered and  the  experiment  is
repeated 10\,000  times.  In  this exercise, we  first use  the entire
optical galaxy catalogue to draw random subsamples without applying an
colour or  optical luminosity cuts. In  99 per cent of  the trials the
number of group  members is lower than the observed  one for the X-ray
AGN sample. Compared to the full optical galaxy population, the excess
of X-ray AGN in groups at $z\approx 1$ is therefore significant at the
99 per  cent level, or  $\approx 2.6 \sigma$  in the case of  a normal
distribution.    This   excess   is   graphically  shown   in   Figure
\ref{fig_vdisp}.  Selection effects  introduced by e.g.  the magnitude
limit  of  the  spectroscopic   survey  $R<24.1$\,mag,  or  the  group
identification method,  are affecting both  the X-ray and  the optical
galaxy  samples  in  the  same   way  and  hence  cancel  out  in  the
differential test used here.

The  environment of  galaxies is  a strong  function of  their optical
colour and luminosity (e.g.  Cooper et al.  2006; Gerke et al.  2007).
X-ray selected AGN  have been shown to be  associated with red (Nandra
et al.   2007) early-type  (Pierce et al.   2007) galaxies,  which are
known  to  be  more  clustered  than blue  sources.   This  effect  is
accounted for by repeating the random resampling of the optical galaxy
population,  after  matching in  each  trial  the absolute  magnitude,
$M_B$, and  the rest-frame $U-B$  colour distributions of  the optical
galaxy subsample  to that of X-ray  sources using bin  sizes of 1\,mag
and 0.5\,mag  respectively.  The redshift distributions  of the random
subsamples and  the X-ray AGN  that are also  statistically identical.
This exercise shows that in 91  per cent of the trials the fraction of
group galaxies is smaller than X-ray AGN, reducing the significance of
the excess  to about $1.7\sigma$.   The group fraction of  the optical
galaxy subsample  with $U-B$ and $M_B$ distributions  matched to those
of X-ray  sources is 33.8 per  cent (see Table 1).  In conclusion, the
higher incidence of  X-ray AGN in groups compared  to optical galaxies
at $z\approx  1$ is to some level  because they are hosted  by red and
luminous  galaxies, which  are  known to  reside  in relatively  dense
regions (Cooper et al. 2006; Gerke et al. 2007).

Contamination  of the optical  light by  emission from  the AGN  is an
issue when matching  the $U-B$ and $M_B$ of  X-ray sources and optical
galaxies. This effect  is likely to be small  however, and is unlikely
to change the results presented here. The majority of X-ray sources in
the    sample    have     $L_X<10^{44}    \rm    erg\,s^{-1}$    (Fig.
\ref{fig_lxdisp}), and  therefore the  central engine is  not powerful
enough to significantly  affect the optical light of  the host galaxy.
For example Nandra et al.  (2007) find only a weak trend between X-ray
luminosity and $M_B$ for the  AEGIS X-ray sources, suggesting that the
optical  light   of  these  systems   is  dominated  by   the  galaxy.
Nevertheless,   AGN   contamination  means   that   the  true   colour
(i.e. corrected  for contamination) of  some AGN hosts is  redder than
observed.  As the fraction of  galaxies in groups increases with $U-B$
colour (Gerke  et al.  2007), it  is expected that  any differences in
the  group fraction  between X-ray  AGN and  optical galaxies  will be
smaller than what is estimated above (e.g.  91 per cent) once the true
(i.e.  corrected  for AGN  contamination) $U-B$ and  $M_B$ of  the two
populations are matched.

\subsection{AGN fraction: groups vs field}

The results of the previous section also indicate that the fraction of
X-ray AGN  relative to optical galaxies (with  matched $U-B$/$M_B$) in
groups and  in the field  differ at the  91 per cent  confidence level
only.  Any differences  in the AGN fraction between  the field and the
group galaxy populations are not large enough to be detected at a high
significance level in the sample presented in this paper.

The  resampling  technique  of  the  previous section  is  useful  for
exploring differences in the fraction of AGN in groups relative to the
field  but cannot  be used  to  estimate these  fractions in  absolute
terms.  Next, we address this  issue to allow comparison with previous
studies at lower redshift.  For this exercise any observational biases
need to be taken into  account explicitly by applying appropriate cuts
to the galaxy  and AGN samples.  The disadvantage  of this approach is
that  the sample  size  is reduced  drastically  (see below),  thereby
increasing the statistical errors of individual measurements.

Firstly,  the  magnitude  limit  of the  DEEP2  spectroscopic  survey,
$R_{AB}<24.1$\,mag, introduces colour  dependent incompleteness in the
sample.  Galaxies with intrinsically red colours, which are on average
more clustered,  drop below the  survey limit at lower  redshifts than
intrinsically  bluer galaxies,  which  are frequently  found in  lower
density regions.   This effect becomes  increasingly severe at  $z \ga
1$, as the $R$-band straddles  the rest-frame UV. Before comparing the
fraction of AGN in groups and in the field, it is essential to account
for this  effect by defining samples  that are complete  for both blue
and red galaxies.   For this exercise we select  galaxies with $z<0.9$
and $M_B<-20$\,mag.  These  cuts result in a sample  that is free from
colour dependent  biases (Gerke et  al.  2007).  The  second selection
effect is related  to the fact that optically  faint group members may
drop out of the spectroscopic  sample at higher redshift.  As a result
a group with certain properties, such as number and optical luminosity
of members, may  be missed by the algorithm at  higher $z$ although it
can be identified  at lower $z$.  To avoid this  bias, we follow Gerke
et al.  (2007) and redefine the group sample to be only those galaxies
that  reside  in  groups  with  two  or  more  members  brighter  than
$M_B=-20$\,mag. In that  way the group sample is  uniform, in terms of
sensitivity to groups of certain properties, across the redshift range
of the sample. For this new sample we find that the fractions of X-ray
AGN in  groups and in the  field are $4.7\pm1.6$ per  cent (8/168) and
$4.5\pm1.0$ per cent (19/425)  respectively, with the errors estimated
using binomial  statistics. The fractions above  are consistent within
the  uncertainties.    X-ray  AGN  are  therefore   found  in  diverse
environments,  from   groups  to  the  field.   Therefore,  the  group
environment  does not  appear to  promote AGN  activity more  than the
field,  once the  properties  (e.g. $M_B/U-B$)  of  the population  of
potential AGN hosts are accounted for.

The small size of the sample is an issue however, and may drown subtle
differences in the AGN fraction between groups and the field. Assuming
that  the  true AGN  fraction  is higher  than  the  observed one,  we
estimate that the  binomial probability of observing exactly  8 AGN in
groups is less  than $<0.3$ per cent for a true  group AGN fraction of
$>10$ per cent. In other words, the sample size used here can only put
an upper limit  of about 10 per cent (99.7  per cent confidence level)
in the  true fraction of  AGN in groups.  Finally, we note  in passing
that our estimate of the fraction of AGN in groups at $z\approx1$ (4.7
per cent, $M_B<-20$\,mag) is similar  to that observed in low redshift
clusters (Martini et al.  2007; 5 per cent, $M_R<-21$\,mag) and groups
(Shen et al.   2007; 7 per cent, $M_R<-21$\,mag)  at a similar optical
luminosity limit.

\subsection{AGN X-ray luminosity distribution: groups vs field}

\begin{figure}
\begin{center}
 \rotatebox{0}{\includegraphics[height=0.9\columnwidth]{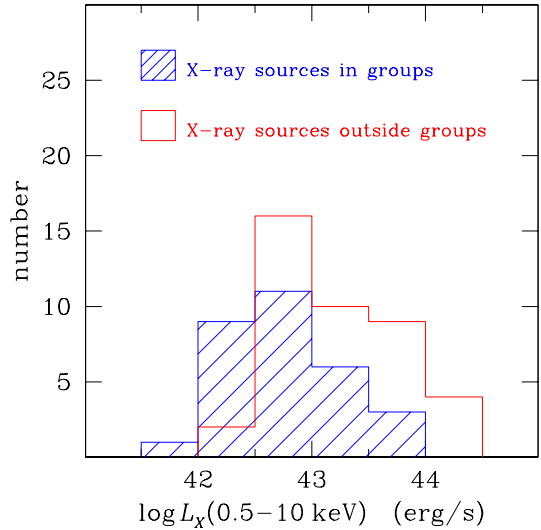}}
\end{center}
\caption{ X-ray luminosity distribution. AEGIS X-ray sources in groups
  correspond to the open (red)  histogram. X-ray sources in the field
  are shown with the hatched (blue)
  histogram. 
}\label{fig_lxdisp}
\end{figure}

Figure \ref{fig_lxdisp} compares the X-ray luminosity distributions of
AGN in groups with those in the field. The latter population is offset
to higher X-ray luminosities.   A Kolmogorov-Smirnov (K-S) tests shows
that  the   likelihood  of  the  observed  differences,   if  the  two
populations  were drawn  from the  same  parent sample,  is $5  \times
10^{-3}$.   However, this  may be  a selection  effect related  to the
group  finding  algorithm.  For  example,  group  galaxies with  given
optical  absolute  magnitudes may  be  identified  as  members at  low
redshift, but at higher redshift  some of them may become fainter than
the spectroscopic magnitude limit  of the AEGIS survey, $R<24.1$\,mag.
In  this case  the group  may not  be identified  by the  algorithm at
higher redshift.   One way to  account for this  bias is to  adopt the
redshift  and optical luminosity  cuts of  the previous  section. This
approach  however,  also significantly  reduces  the  sample size  and
increases  the random  errors. Instead,  we  use the  full sample  and
account  for redshift  dependent  selection effects  using the  random
resamples  of  the optical  galaxy  population,  after matching  their
absolute magnitude  and optical colour distributions to  that of X-ray
sources.  Each  galaxy in individual  subsamples is assigned  an X-ray
flux, drawn randomly and  without replacement from the observed fluxes
of the AEGIS X-ray sources.   This results in subsamples with the same
flux distribution as  X-ray AGN.  The K-S test is  then applied to the
X-ray luminosities  of the mock subsample to  estimate the probability
that  group members  and non-group  members  are drawn  from the  same
parent population.  It  is found that for the  mock catalogues the K-S
likelihood  of the  observed differences  in $L_X$  between  group and
non-group members,  if the  two populations were  drawn from  the same
parent population, is larger than the observed one for X-ray AGN in 98
per cent  of the trials.   This is tentative evidence,  significant at
the 98  per cent  level, that  at $z\approx 1$  X-ray selected  AGN in
groups are systematically less X-ray luminous compared to those in the
field.

\subsection{AGN host galaxy properties}

\begin{figure}
\begin{center}
 \rotatebox{0}{\includegraphics[height=0.9\columnwidth]{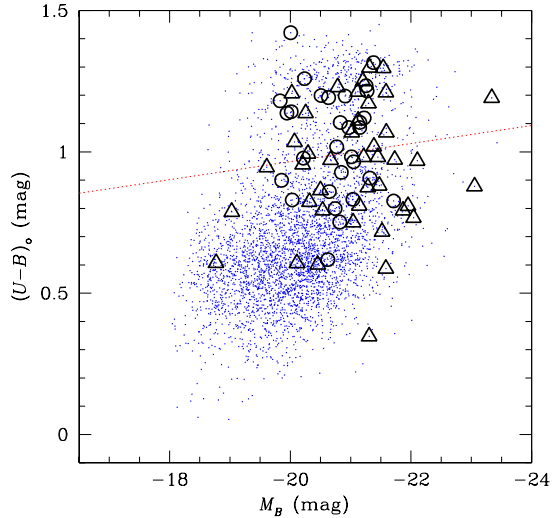}}
\end{center}
\caption{Rest-frame $U -  B$ colour against B-band absolute magnitude
  for DEEP2  galaxies (blue dots)  and X-ray AGN in groups (black
  circles) and outside groups (black  triangles). All sources are in
  the redshift range $0.7<z<1.4$. The diagonal dotted line  is
  defined by Willmer et   al.  (2006)   to separate   the  red   from
  the   blue  clouds.  
}\label{fig_cmd}
\end{figure}

\begin{figure}
\begin{center}
 \rotatebox{0}{\includegraphics[height=0.9\columnwidth]{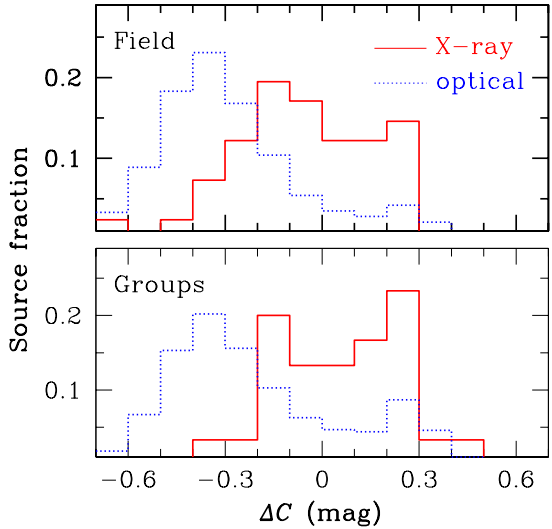}}
\end{center}
\caption{ Distribution of the $\Delta  C$ parameter for X-ray AGN (red
continuous  line)  and optical  galaxies  (dotted  blue histogram)  in
groups (bottom panel) and in the field (upper panel).  $\Delta C$ is a
measure of  the position of  a source on the  colour-magnitude diagram
(Georgakakis et al. 2008) and is defined as the difference between the
colour of the  galaxy and the line $U-B=  -0.032 \, (M_B+21.52)+1.014$
(AB  units), which  separates  the blue  from  the red  clouds in  the
$U-B$/$M_B$ colour-magnitude diagram  at $z\approx1$ (Willmer  et al.
2006). The $\Delta C$ bins plotted  here are parallel to this line and
their position in  the CMD depends on $M_B$.  $\Delta C=0$ corresponds
to the valley between the blue and the red clouds, $\Delta C>0$ is for
red  cloud galaxies  and  $\Delta  C<0$ is  for  blue cloud  galaxies.
}\label{fig_dc}
\end{figure}

The colour-magnitude  diagram is used to  explore possibly differences
in  the host  galaxy properties  of AGN  in groups  and in  the field.
Figure \ref{fig_cmd} shows the distribution of the group and the field
X-ray AGN  in the $U-B$ vs $M_B$  colour-magnitude diagram.  Optically
luminous AGN ($M_B\la-21$\,mag) are systematically found in the field,
in  agreement with  the  results presented  in  the previous  section.
Nevertheless,  both   the  field  and  the  group   AGN  have  similar
distributions in the $U-B$ colour.  This is further demonstrated using
the parameter $\Delta  C$ (Georgakakis et al. 2008),  which is defined
as  the difference  between  the colour  of  the galaxy  and the  line
separating the blue from the red clouds (Willmer et al.  2006). Figure
\ref{fig_dc} plots  the distribution of  the $\Delta C$  parameter for
X-ray sources  and optical  galaxies in groups  and in the  field.  In
both these environments AGN  have similar distributions and occupy the
region of the  colour magnitude space between the  red and blue clouds
(Nandra et al.  2007;  Georgakakis et al. 2008).  A Kolmogorov-Smirnov
tests shows  that the  likelihood of the  observed differences  in the
$\Delta C$  distributions if  the field and  the group AGN  were drawn
from  the same parent  population is  27 per  cent.  We  conclude that
there is no evidence for a  difference in the rest-frame colour of AGN
hosts in the field and in groups.  This contrary to recent claims that
AGN associated  with the  large scale structures  in the  Chandra Deep
Field  South (Gilli  et al.   2003)  are preferentially  found in  the
valley between  the blue  and the red  clouds of  the colour-magnitude
distribution of galaxies (Silverman et al.  2007).

\section{Discussion}

Using data from  the AEGIS survey we explore, for  the first time, the
incidence  of active  SBHs in  groups at  high  redshift, $z\approx1$.
X-ray  selected AGN at  this redshift  are more  frequently associated
with groups compared  to the overall optical galaxy  population at the
99  per  cent  confidence  level.   This  result  is  consistent  with
independent estimates of the large  scale distribution of X-ray AGN at
$z\approx1$ using  either the auto-correlation  function (e.g.  Miyaji
et al.   2007), measurements of the  local density in  the vicinity of
X-ray  sources (Georgakakis  et al.   2007), or  the cross-correlation
function with  optical galaxies (Coil  et al.  2008 in  prep.).  These
studies  find  that  the  typical  environment  of  X-ray  sources  at
$z\approx1$  is  similar to  or  possibly  even  denser than  that  of
early-type galaxies,  which are  known to be  more clustered  than the
overall galaxy population (e.g.  Coil et al.  2008).  Given that X-ray
AGN  at  $z\approx1$ are  hosted  mostly  by  red early-type  galaxies
(Pierce et al.  2007; Nandra et  al.  2007), it is not surprising that
they  are  preferentially  found   in  high  density  regions.   After
accounting for the colour and optical luminosity distribution of X-ray
AGN hosts,  it is found that the  excess of AGN in  groups compared to
galaxies of similar intrinsic optical properties is significant at the
91 per cent only.  This is  consistent with the work of Georgakakis et
al.   (2007)  who  found  no statistically  significant  environmental
differences  between $z\approx1$  X-ray  AGN and  galaxies of  similar
rest-frame optical  colour and luminosity.  Low  redshift studies also
report similar  results.  At $z\approx0.1$, the  environment of narrow
optical emission line  AGN on large scales is almost  the same as that
of non-AGN with similar host galaxy properties (Li et al.  2006).

Another  way  of  presenting  the  results  above  is  that  once  the
properties of  AGN hosts  are accounted for,  the group and  the field
environment at $z\approx1$ are different at the 91 per cent confidence
level in the fraction of X-ray AGN relative to galaxies.  Although the
small number of sources is a  problem, this result suggests a range of
environments for X-ray  AGN at $z\approx1$. At low  redshift, there is
also no strong evidence for a significant variation in the fraction of
AGN  relative  to  galaxies  at different  environments.   Popesso  \&
Biviano (2006)  showed that the fraction of  optically selected narrow
emission-line  AGN in  low redshift  clusters anticorrelates  with the
velocity dispersion  and becomes consistent  with the field  value for
densities  similar to  or below  those  of groups  and poor  clusters.
Martini et al.  (2006, 2007) however, argue that optical AGN selection
methods are missing  a substantial fraction of AGN  in clusters, which
can  be identified by  X-ray observations.   These authors  argue that
when  the  X-ray AGN  are  taken into  account,  there  are no  strong
differences  between  the field  and  clusters  either.   Shen et  al.
(2007) using a small sample  of groups at $z\approx0.1$ found that the
fraction of both  optically and X-ray selected AGN  is consistent with
the fraction  of X-ray AGN in  clusters (e.g.  Martini  et al.  2007).
Although  the  studies  above   are  still  limited  by  small  number
statistics, that may drown weak  trends, and also by inhomogeneous AGN
selection  criteria (i.e.   X-ray vs  optical  narrow emission-lines),
they  suggest that  AGN activity  does  not strongly  depend on  local
density.  Therefore,  the mechanism(s) responsible  for triggering the
accretion on SBHs operate nearly  equally efficiently, in terms of AGN
numbers,  in groups, the  field and  possibly in  clusters. It  is the
preference of AGN  for luminous/red hosts that makes  them appear more
clustered than the overall  galaxy population. Recent simulations also
suggest  diverse environments for  AGN.  Colberg  \& Di  Matteo (2008)
showed that  the most  massive and active  SBHs at $z\approx1$  are in
groups,  albeit with  significant scatter,  while  less massive/active
SBHs  span  a  wide range  of  local  densities,  from groups  to  the
field. The  X-ray AGN  sample used here  is likely to  include systems
with a wide range of BH masses and accretion rates (e.g.  Bundy et al.
2008).  It is therefore not  surprising that we find similar fractions
of AGN in groups and in the field.

Although the fraction of  AGN at different environments is comparable,
there is  tentative evidence, significant  at 98 per cent  level, that
the field  can produce more powerful  AGN compared to  groups. Coil et
al. (2007) also found that luminous broad-line QSOs are clustered like
blue galaxies  suggesting low density environments  for these systems.
The results  above are at odds  with the simulations of  Colberg \& Di
Matteo (2008) at $z=1$ and with recent observations which suggest that
the  clustering of X-ray  selected AGN  at $z\approx1$  increases with
X-ray luminosity (Plionis et  al.  2008).  This discrepancy underlines
the  need to  further  explore  the dependence  of  AGN clustering  on
luminosity.    Interestingly  however,   low   redshift  studies   are
consistent with  our finding.  Low  luminosity AGN (e.g.   LINERs) are
more clustered than the  overall AGN population at $z\approx0.1$ (Wake
et al.  2004;  Constantin \& Vogeley 2006).  Kauffmann  et al.  (2003)
also  found  a strong  environmental  dependence  of  the fraction  of
luminous ($L_{\rm  [O\,III]\, 5007}  \rm > 10^{7}  \, erg  \, s^{-1}$)
optically selected (narrow-line)  AGN at $z\approx0.1$.  The incidence
of such objects increases with decreasing local density, from clusters
to the field.  Popesso \& Biviano (2006) found a similar trend for the
mean equivalent width of the [O\,II]\,3727 emission line of AGN in low
redshift  clusters.   Active SBHs  with  higher  equivalent widths  of
[O\,II]\,3727  reside,  on   average,  in  lower  velocity  dispersion
clusters.   Although the  Popesso \&  Biviano (2006)  sample  does not
include very  low density  environments, such as  small groups  or the
field, extrapolation of their best-fit relations below $\rm \sigma_v =
200 \, km  \, s^{-1}$, implies that the fraction  of luminous AGN will
continue  to increase  with decreasing  density.   An anti-correlation
between AGN luminosity and  local density at $z\approx1$, if confirmed
with a larger sample, may  be related to either gas availability (i.e.
galaxies  in groups  have less  fuel  to feed  the supermassive  black
hole), or the nature of the gravitational interactions in groups (i.e.
prolonged tidal  disruptions, minor mergers), or a  combination of the
two.

In addition  to possible differences  in the X-ray luminosity  and the
number  of  AGN,  we  also  explore  evidence  that  the  host  galaxy
properties  of active SBH  in groups  and in  the field  are distinct.
Silverman et  al.  (2007) for  example, found that the  fraction X-ray
AGN  in  the  valley between  the  blue  and  the  red clouds  of  the
colour-magnitude diagram  increases in  the large scale  structures of
the Chandra Deep Field South,  identified by the redshift spikes at $z
= 0.67$  and $z =  0.73$ in that  field (Gilli et al.   2003).  Figure
\ref{fig_dc} shows that  AGN host galaxies in the  field and in groups
have similar distributions in the colour-magnitude diagram. Both these
populations  are associated with  galaxies in  the valley  between the
blue and the red clouds.  The  data does not support an association of
AGN host galaxy  colour and large scale structures  like that found by
Silverman  et al.   (2007).  We  note  however, that  the ``wall''  or
``sheet'' structures  in the Chandra  Deep Field South,  identified as
spikes in  the redshift distribution,  are likely to  be significantly
different in terms of spatial extent and virialisation state, compared
to the groups studied here.

\section{Conclusions}

In  conclusion,  it  is  found  that  at  $z\approx1$,  AGN  are  more
frequently found in groups  compared to the overall optically selected
galaxy population at the 99  per cent confidence level. This excess is
to some level  because X-ray AGN are associated  with red and luminous
galaxies which are known to  be more clustered than the overall galaxy
population.  After taking into  account the AGN host galaxy properties
the significance of the excess of  AGN in groups drops to 91 per cent.
This suggests that X-ray AGN live  in a range of environments and that
groups do not produce more AGN  than the field, once their host galaxy
properties are factored in  the analysis.  There is tentative evidence
that AGN in the field are  more powerful, on average, than their group
counterparts,   extending   similar  results   at   low  redshift   to
$z\approx1$.  This can be attributed to differences in the gas content
of  group/field galaxies and/or  the nature  of interactions  in these
environments.

Although  the results  presented  here are  affected  by small  number
statistics, it  is the first time  that an attempt is  made to explore
the relation between SBH  accretion and optically identified groups at
$z\approx1$.  There are two obvious ways to improve the reliability of
our  conclusions.  Either  perform deep  X-ray observations  in fields
with extensive optical spectroscopy,  like those surveyed by the DEEP2
project,  or complement  existing  deep X-ray  surveys (e.g.   COSMOS,
Chandra Deep Fields) with  optical spectroscopy of similar quality and
extent to the DEEP2 spectroscopic survey.

\section{Acknowledgments} 

The  authors  wish  to  thank  the  anonymous  referee  for  providing
constructive comments and suggestions that significantly improved this
paper.  This work  has been supported by funding  from the Marie-Curie
Fellowship  grant  MEIF-CT-2005-025108 (AG)  and  STFC  (ESL). BFG  is
supported  by the  U.S.  Department  of Energy  under  contract number
DE-AC02-76SF00515.  ALC is supported by NASA through Hubble Fellowship
grant HF-01182.01-A, awarded by the Space Telescope Science Institute,
which is operated  by the Association of Universities  for Research in
Astronomy, Inc., for NASA, under contract NAS 5-26555.

\end{document}